\documentclass[aps,prd,preprint,groupedaddress,preprintnumbers]{revtex4-1}

\usepackage{graphicx}
\newcommand{\nn}{\nonumber}
\newcommand{\e}{\hbox{e}}
\newcommand{\dt}{\frac{d}{dt}}
\newcommand{\xbar}{\bar{x}}
\newcommand{\mbar}{\overline{m^2}}
\newcommand{\xph}{x_{\mathrm{ph}}}
\newcommand{\tph}{t_{\mathrm{ph}}}
\newcommand{\mph}{m^2_{\mathrm{ph}}}
\newcommand{\lb}{\left\lbrace}
\newcommand{\rb}{\right\rbrace}
\newcommand{\Veff}{V_{\mathrm{eff}}}
\begin{document}

\preprint{KOBE-TH-13-05}

\title{Solving RG equations with the Lambert $W$ function}


\author{H.~Sonoda}
\email[]{hsonoda@kobe-u.ac.jp}
\affiliation{Physics Department, Kobe University, Kobe 657-8501 Japan}


\date{26 March 2013}

\begin{abstract}
  It has been known for some time that 2-loop renormalization group
  (RG) equations of a dimensionless parameter can be solved in a
  closed form in terms of the Lambert $W$ function.  We apply the
  method to a generic theory with a Gaussian fixed point to construct
  RG invariant physical parameters such as a coupling constant and a
  physical squared mass.  As a further application, we speculate a
  possible exact effective potential for the $O(N)$ linear sigma model
  in four dimensions.
\end{abstract}

\pacs{11.10.Gh, 11.10.Hi, 11.15.Bt}

\maketitle

\section{Introduction}

The purpose of this paper is to solve generic 2-loop renormalization
group (RG) equations exactly using the Lambert $W$
function.\cite{wiki:LambertW} The Lambert $W$ function has been
introduced previously to solve analytically the 2- and 3-loop RG
equations for QCD.\cite{Gardi:1998qr, Magradze:1998ng,
  Magradze:1999um} (See also \cite{Nesterenko:2003xb} for a review of
various applications of the Lambert $W$ function to QCD.)  We apply
the same function to solve generic 2-loop RG equations for theories
with a Gaussian fixed point such as the $O(N)$ non-linear sigma model
in four dimensions.  (This has been partially done in
\cite{Curtright:2010hq}.) 

In the following we wish to justify our purpose by reminding the
reader of the generality of 2-loop RG equations.
We consider two examples in four dimensions: QCD and the $\phi^4$
theory.  Let us consider QCD first.  

Let $\Lambda_0$ be the ultraviolet cutoff, and $g_0^2$ be the bare
gauge coupling, normalized appropriately.  To construct the continuum
limit ($\Lambda_0 \to \infty$) we must give a particular $\Lambda_0$
dependence to $g_0^2$:
\begin{equation}
g_0^2 = \frac{1}{\ln \frac{\Lambda_0}{\mu} + c \ln \ln
  \frac{\Lambda_0}{\mu} - \ln \frac{\Lambda (g^2)}{\mu}}
\end{equation}
where $c = \frac{6 \times 153}{33}$ for QCD with no quarks.  We can
introduce a gauge coupling $g^2$, renormalized at a renormalization
scale $\mu$, through the $\Lambda_0$ independent constant in the
denominator.  If we choose
\begin{equation}
\frac{\Lambda (g^2)}{\mu} = \e^{- \frac{1}{g^2}} \left( \frac{1}{g^2}
    + c \right)^c
\end{equation}
then $g^2$ satisfies the 2-loop RG equation
\begin{equation}
- \mu \frac{\partial}{\partial \mu} g^2 = (g^2)^2 + c (g^2)^3
\end{equation}
exactly.

We next consider the $\phi^4$ theory defined by the bare action
\begin{equation}
S = \int d^4 x\, \left( \frac{1}{2} \partial_\mu \phi \partial_\mu
  \phi + \frac{m_0^2}{2} \phi^2 + \frac{\lambda_0}{4!} \phi^4 \right)
\end{equation}
with an ultraviolet cutoff $\Lambda_0$.  This is a theory with the
Gaussian fixed point $m_0^2 = \lambda_0 = 0$.  We cannot take
$\Lambda_0$ all the way to infinity, but for $\Lambda_0$ large
compared with the physical mass, we obtain an almost continuum limit.
For a given $\lambda_0$, let the critical squared mass be
\begin{equation}
m_{\mathrm{cr}}^2 (\lambda_0) = A_4 (\lambda_0) \Lambda_0^2
\end{equation}
Then, to get an almost continuum limit, we tune the bare squared mass
as
\begin{equation}
m_0^2 = A_4 (\lambda_0) \Lambda_0^2 + z_m (\lambda_0)
  \left(\frac{\frac{(4\pi)^2}{3 \lambda_0} - c}{\frac{(4\pi)^2}{3
        \lambda} - c}\right)^a m^2 
\end{equation}
where
\begin{equation}
c = \frac{17}{27},\quad a = - \frac{1}{3}
\end{equation}
Here, the renormalized coupling $\lambda$ is defined so that
\begin{equation}
  \frac{\Lambda_0}{\mu} =  \e^{t_0 (\lambda_0) + \frac{(4 \pi)^2}{3 \lambda} -
    \frac{(4 \pi)^2}{3 \lambda_0}} 
  \left( \frac{\frac{(4\pi)^2}{3 \lambda} - c}{\frac{(4 \pi)^2}{3
        \lambda_0} - c} \right)^c 
\end{equation}
The $\lambda_0$ dependence of $z_m$ and $t_0$ is determined so that
the theory with given $m^2$ and $\lambda$ have no $\lambda_0$
dependence except for non-universal contributions suppressed by
inverse powers of $\Lambda_0$.  The parameters $\lambda, m^2$
renormalized at the scale $\mu$ satisfy the 2-loop RG equations
(1-loop for $m^2$) exactly:
\begin{equation}
\lb\begin{array}{c@{~=~}l}
- \mu \frac{\partial}{\partial \mu} \frac{3 \lambda}{(4 \pi)^2} & -
\left(\frac{3 \lambda}{(4\pi)^2}\right)^2 + c \left(\frac{3
      \lambda}{(4\pi)^2}\right)^3\\
- \mu \frac{\partial}{\partial \mu} m^2 & a \frac{3
      \lambda}{(4\pi)^2} m^2
\end{array}\right.
\label{generic}
\end{equation}

We have thus reminded the reader that renormalization schemes exist so
that 2-loop RG equations become exact.  Hence, solving 2-loop RG
equations amounts to solving general RG equations.  This paper is
organized as follows.  In sect.~\ref{RGeqs} we solve the generic
2-loop RG equations (\ref{generic}) exactly in terms of the Lambert
$W$ function.  (This has actually been done already in sect.~II.B.1 of
\cite{Curtright:2010hq}.)  Then, in sec.~\ref{physical}, we give the
main results of this paper by constructing two physical parameters:
one corresponding to the dimensionless coupling and the other
corresponding to a physical squared mass.  In sect.~\ref{inversion} we
invert the construction and express the renormalized parameters in
terms of the physical parameters.  In sect.~\ref{Veff}, we generalize
the exact effective potential in the large $N$ limit of the $O(N)$
linear sigma model \cite{Sonoda:2013a} to construct a trial effective
potential for finite $N$, fully consistent with the 2-loop RG
equations.

In this paper we adopt the convention to fix the renormalization scale
at $\mu = 1$.  Hence, a squared mass parameter acquires the canonical
dimension $2$ in addition to the anomalous dimension in its RG
equation.

\section{Generic 2-loop RG equations\label{RGeqs}}

We consider the following generic 2-loop RG equation:
\begin{equation}
\dt x = - x^2 + c x^3
\end{equation}
and
\begin{equation}
\dt m^2 = (2 + a x) m^2
\end{equation}
For example, in the $O(N)$ linear sigma model in four dimensions, we
find
\begin{equation}
c = \frac{9 N + 42}{(N+8)^2}
\label{c-ON}
\end{equation}
for the self-coupling, and
\begin{equation}
a = - \frac{N+2}{N+8}
\label{a-ON}
\end{equation}
for the squared mass parameter.

In the following we assume $c > 0$ and $0 \le x \ll \frac{1}{c}$.
Let us define a mass scale by
\begin{equation}
\Lambda (x) \equiv \lb \e^{\frac{1}{c x} - 1} \left(\frac{1}{c x} -
      1 \right)\rb^c \gg 1
\end{equation}
This satisfies
\begin{equation}
\dt \Lambda (x) = \Lambda (x)
\end{equation}
$\Lambda (x)$ is of the same order as the UV cutoff.  We can invert
the definition of $\Lambda (x)$ to express $x$ in terms of $\Lambda
(x)$.  Since
\begin{equation}
\Lambda (x)^{\frac{1}{c}} = \e^{\frac{1}{c x}-1} \left( \frac{1}{c x}
    - 1 \right)
\end{equation}
we obtain
\begin{equation}
\frac{1}{c x} - 1 = W \left( \Lambda (x)^{\frac{1}{c}} \right)
\end{equation}
where $W$ is the upper branch of the Lambert $W$ function defined by
\begin{equation}
W (x) \e^{W (x)} = x
\end{equation}
for $x \ge - \frac{1}{\e}$. (See Appendix 1.)

We now define the running parameter $\xbar (t; x)$ by
\begin{equation}
\frac{1}{c \xbar (t;x)} - 1 = W \left( \left(\e^t \Lambda
        (x)\right)^{\frac{1}{c}} \right)
\end{equation}
or equivalently by
\begin{equation}
\xbar (t;x) = \frac{1}{c \left(1 + W \left( \left(\e^t \Lambda
        (x)\right)^{\frac{1}{c}} \right)\right)}
\label{xbar}
\end{equation}
so that it satisfies both
\begin{equation}
\partial_t \xbar (t;x) = - \xbar (t;x)^2 + c \xbar (t;x)^3
\end{equation}
and the initial condition
\begin{equation}
\xbar (0;x) = x
\end{equation}
(Eq.~(\ref{xbar}) agrees with (26) of \cite{Curtright:2010hq} which
gives the same result for $c = 1$.)

Analogously, the running parameter $\mbar (t;x,m^2)$ defined by
\begin{equation}
\mbar (t;x, m^2) \equiv \e^{2 t} m^2 \left( \frac{\frac{1}{c \xbar (t;x)} -
    1}{\frac{1}{c x} - 1} \right)^a
\end{equation}
satisfies
\begin{equation}
\partial_t \mbar (t; x, m^2) = \left(2 + a \xbar (t;x) \right) \mbar
(t; x, m^2)
\end{equation}
and the initial condition
\begin{equation}
\mbar (0; x, m^2) = m^2
\end{equation}

\section{Physical parameters\label{physical}}

We now apply the results of the previous section to construct physical
parameters.  We first observe that the combination
\begin{equation}
\frac{m^2}{\left(\frac{1}{c x} - 1\right)^a \Lambda (x)^2}
= m^2 \frac{\e^{- 2 c \left(\frac{1}{c x} - 1\right)}}{\left( \frac{1}{c
      x} - 1 \right)^{a + 2 c}} 
\end{equation}
is an RG invariant.  We then note that any RG invariant of $x$ and
$m^2$ can be obtained as a function of the above RG invariant.

Let $\xph (x, m^2)$ be the RG invariant satisfying
\begin{equation}
\xph (x, 1) = x \label{xph-initial}
\end{equation}
To obtain an explicit expression for $\xph (x, m^2)$, let us write it
in the form
\begin{equation}
\xph (x, m^2) = f \left(  (m^2)^{- \frac{1}{a + 2 c}} \frac{2 c}{a + 2
    c} \left(\frac{1}{c x} - 1 \right) \e^{\frac{2 c}{a + 2 c}
    \left(\frac{1}{c x} - 1\right)} \right)
\end{equation}
where we assume $m^2 > 0$.  The condition (\ref{xph-initial}) implies
\begin{equation}
f \left( \frac{2 c}{a + 2
    c} \left(\frac{1}{c x} - 1 \right) \e^{\frac{2 c}{a + 2 c}
    \left(\frac{1}{c x} - 1\right)} \right) = x
\end{equation}
We can rewrite this as
\begin{equation}
\frac{2 c}{a + 2 c} \left( \frac{1}{c f (s \e^s)} - 1 \right)
= s \equiv \frac{2 c}{a + 2 c} \left( \frac{1}{c x} - 1\right)
\end{equation}
For small $x \ll 1$, we find $s \gg 1$ for $a + 2 c > 0$, and $- s \gg
1$ for $a + 2 c < 0$.  The above equation is solved by the Lambert $W$
function as
\begin{equation}
\frac{2 c}{a + 2 c} \left( \frac{1}{c f (s \e^s)} - 1 \right)
= \lb\begin{array}{c@{\quad\textrm{if}\quad}l}
W (s \e^s)& a + 2 c > 0\\
W_{-1} (s \e^s)& a + 2 c < 0
\end{array}\right.
\end{equation}
We thus obtain the physical coupling $\xph (x,m^2)$ as
\begin{equation}
\frac{2 c}{a + 2 c} \left( \frac{1}{c \xph (x,m^2)} - 1 \right)
= \lb\begin{array}{c@{\quad\textrm{if}\quad}l}
W \left((m^2)^{- \frac{1}{a + 2 c}} \frac{2 c}{a + 2 c}
  \left(\frac{1}{c x} - 1\right) \e^{\frac{2c}{a+2c}\left(\frac{1}{cx}
      - 1\right)} \right)& a + 2 c > 0\\
W_{-1} \left( (m^2)^{- \frac{1}{a + 2 c}} \frac{2 c}{a + 2 c}
  \left(\frac{1}{c x} - 1\right) \e^{\frac{2c}{a+2c}\left(\frac{1}{cx}
      - 1\right)} \right)& a + 2 c < 0
\end{array}\right.
\label{xph}
\end{equation}
We plot the left-hand side as a function of $m^2$ for $x = 0.1$
assuming $c = \frac{17}{27}$ and $a = - \frac{1}{3}$.
\begin{figure}[h]
\centering
\includegraphics[width=8cm]{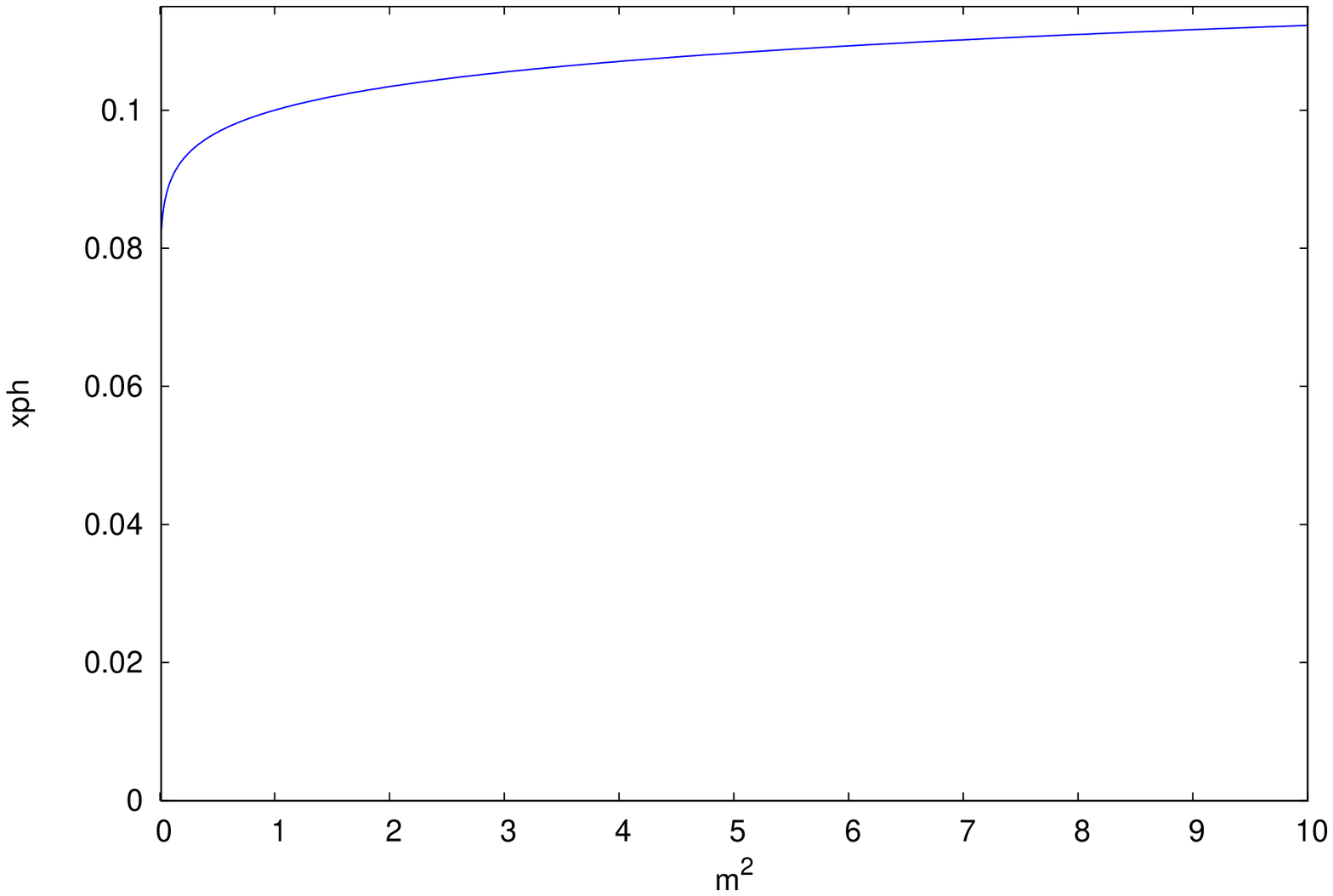}
\includegraphics[width=8cm]{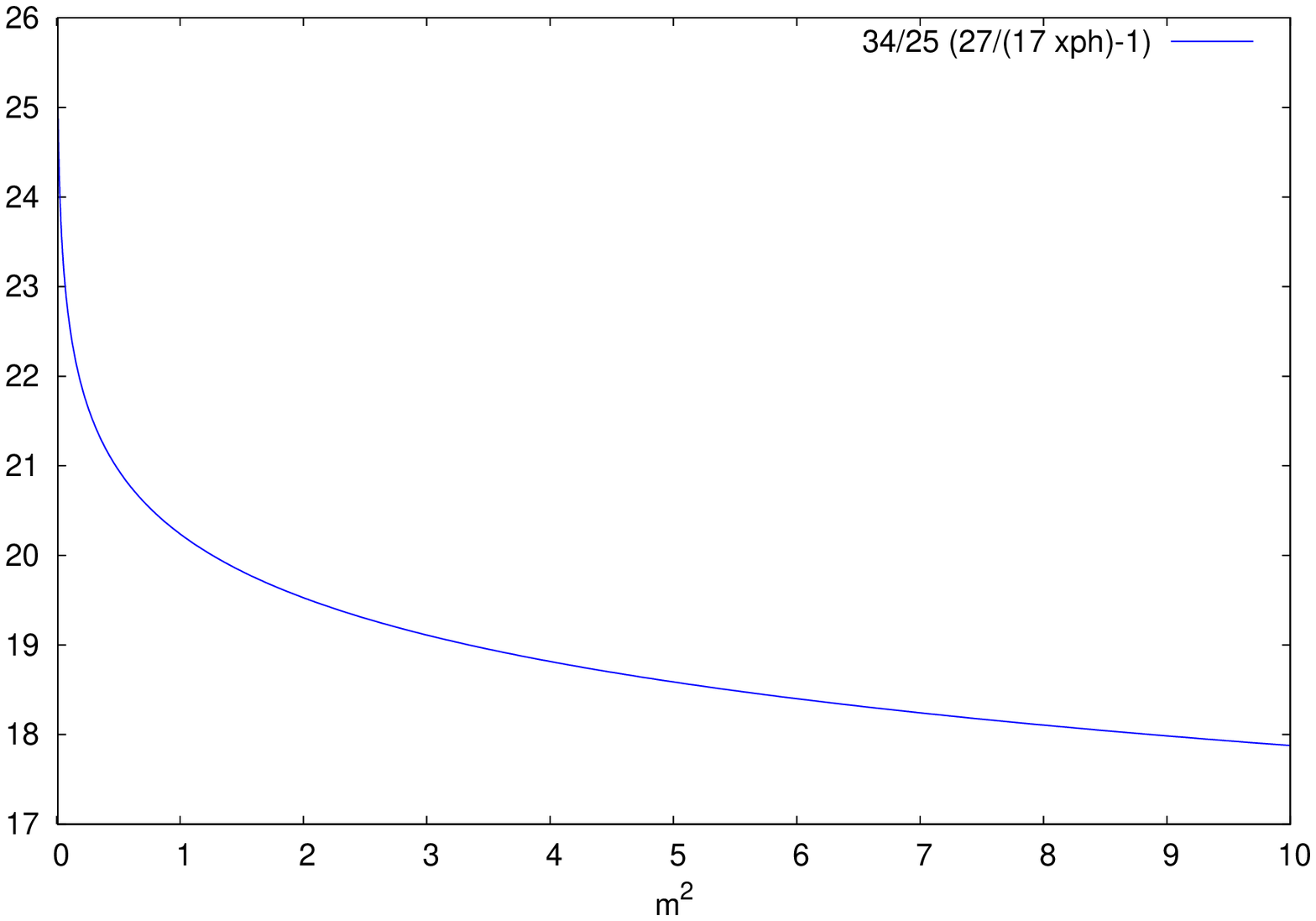}
\caption{Plots of $\xph$ (left) and $\frac{34}{25} \left(\frac{27}{17
        \xph} - 1\right)$ (right) for $x=0.1$, $c = \frac{17}{27}, a =
  - \frac{1}{3}$: $\xph$ increases monotonically as a function of
  $m^2$.  It vanishes as $\frac{1}{- \ln m^2}$ as $m^2 \to 0$, and
  approaches $\frac{27}{17}$ as $m^2 \to \infty$.}
\label{xph-plot}
\end{figure}

Though it is not obvious, the physical coupling $\xph$ admits an
asymptotic expansion in powers of $x$.  This is because $\xph$ can be
defined by the differential equation
\begin{equation}
\dt \xph \equiv \left[ \left(- x^2 + c x^3 \right) \partial_x + (2 + a
  x) m^2 \partial_{m^2} \right] \xph = 0
\label{xph-diffeq}
\end{equation}
and the initial condition (\ref{xph-initial}).  For small $x \ll 1$,
we can expand $\xph$ asymptotically in powers of $x$ in the form
\begin{equation}
\xph (x,m^2) = x \left[ 1 + \sum_{n=1}^\infty x^n p_n (\ln m^2) \right]
\end{equation}
where $p_n$ is a polynomial of degree $n$ satisfying $p_n (0) = 0$.
In principle, this can be shown directly from (\ref{xph}), but it is
more easily shown from the differential equation (\ref{xph-diffeq}).

The physical coupling can also be given in the form of a running
parameter:
\begin{equation}
\xph (x, m^2) = \xbar (- \tph (x,m^2); x)
\end{equation}
where $\tph (x,m^2)$ satisfies
\begin{equation}
\dt \tph (x,m^2) = 1
\end{equation}
and
\begin{equation}
\tph (x, 1) = 0
\end{equation}
To find $\tph (x, m^2)$, we use the defining equality $W (x \e^x) =
x$ to obtain
\begin{eqnarray}
\frac{1}{c \xph (x,m^2)} - 1 &=& W \left( \left(\frac{1}{c \xph
      (x,m^2)} - 1 \right)
\e^{\frac{1}{c \xph (x,m^2)} - 1} \right)\nn\\
&=& W \left( \Lambda (x)^{- \frac{1}{c}} \left(\frac{1}{c \xph
      (x,m^2)} - 1 \right)
\e^{\frac{1}{c \xph (x,m^2)} - 1} \cdot \Lambda (x)^{\frac{1}{c}}
\right)\nn\\
&=& W \left( \e^{- \frac{1}{c} \tph (x,m^2)}  \Lambda
  (x)^{\frac{1}{c}} \right)
\end{eqnarray}
Hence, we obtain
\begin{equation}
\tph (x,m^2) = \ln \Lambda (x) - c \ln \lb\left(\frac{1}{c \xph (x,m^2)}
  - 1 \right) \e^{\frac{1}{c \xph (x,m^2)} - 1} \rb
\end{equation}

In addition to the physical coupling, we can introduce a physical
squared mass by
\begin{equation}
\mph (x,m^2) \equiv m^2 \left( \frac{\frac{1}{c \xph (x,m^2)} -
    1}{\frac{1}{c x} - 1} \right)^a
\label{mph}
\end{equation}
This satisfies
\begin{equation}
\dt \mph (x,m^2) = 2 \mph (x, m^2)
\end{equation}
and the initial condition
\begin{equation}
\mph (x, 1) = 1
\end{equation}
Using (\ref{xph}), we can rewrite the physical squared mass as
\begin{equation}
\mph (x,m^2) = \frac{m^2}{\left(\frac{1}{x} - c\right)^a} \lb
\frac{a + 2 c}{2} W \left( \frac{2 c^{- \frac{2 c}{a+2c}}}{a+2c} \left(
    \frac{m^2}{\left(\frac{1}{x} - c\right)^a\, \Lambda (x)^2}
  \right)^{- \frac{1}{a + 2 c}}\right) \rb^a
\label{mph-expression}
\end{equation}
$W$ should be replaced by $W_{-1}$ if $a + 2 c < 0$.  The physical
squared mass also admits an asymptotic expansion in $x$ just as
$\xph$:
\begin{equation}
\mph (x,m^2) = m^2 \left[ 1 + \sum_{n=1}^\infty x^n q_n (\ln m^2)
\right]
\end{equation}
where $q_n$ is a polynomial of degree $n$ satisfying $q_n (0) = 0$.
\begin{figure}[h]
\centering
\includegraphics[width=8cm]{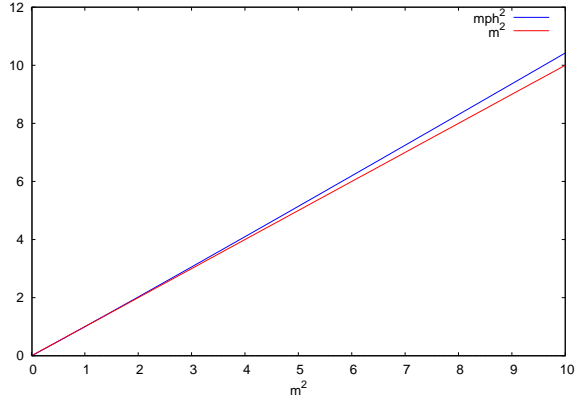}
\caption{Plot of $\mph$ and $m^2$: We plot $\mph$ for $x = 0.1$, $c =
  \frac{17}{27}$, $a = - \frac{1}{3}$. $\mph$ is essentially a
  monotonically increasing function of $m^2$, even though it
  eventually starts decreasing when $m^2$ reaches the UV cutoff scale.}
\label{mph-plot}
\end{figure}

\section{$x$, $m^2$ in terms of $\xph$, $\mph$\label{inversion}}

In the above we have introduced two physical parameters $\xph, \mph$
as functions of $x, m^2$.  We can invert their relations to express
$x, m^2$ in terms of $\xph, \mph$.  We first rewrite (\ref{xph}) and
(\ref{mph}) as
\begin{eqnarray}
\left(\frac{1}{c x} - 1\right) \e^{\frac{2c}{a+2c}
  \left(\frac{1}{c x} - 1\right)} (m^2)^{- \frac{1}{a + 2 c}}
&=& \left(\frac{1}{c \xph} - 1 \right) \e^{\frac{2
    c}{a + 2 c} \left( \frac{1}{c \xph} - 1 \right)}\\
m^2 \left(\frac{1}{cx} - 1 \right)^{-a} &=& \mph \left( \frac{1}{c
    \xph} - 1 \right)^{-a}
\end{eqnarray}
where we assume $m^2 > 0$, and $W$ should be replaced by $W_{-1}$ if
$a + 2 c < 0$.  Substituting the second equation into the first to
eliminate $m^2$, we obtain
\begin{equation}
\left(\frac{1}{c x} - 1 \right) \e^{\frac{1}{c x} - 1}
= (\mph)^{\frac{1}{2 c}} \left( \frac{1}{c \xph} - 1 \right)
\e^{\frac{1}{c \xph} - 1}
\end{equation}
This gives
\begin{equation}
\frac{1}{c x} - 1 = W \left( (\mph)^{\frac{1}{2 c}} \left( \frac{1}{c
      \xph} - 1 \right) \e^{\frac{1}{c \xph} - 1} \right)
\end{equation}
which is valid irrespective of the sign of $a + 2 c$.
Hence, we obtain
\begin{equation}
x = \frac{1}{c} \frac{1}{1 + W \left( (\mph)^{\frac{1}{2 c}} \left( \frac{1}{c
      \xph} - 1 \right) \e^{\frac{1}{c \xph} - 1} \right)}
\end{equation}

Using this result, we then obtain
\begin{eqnarray}
m^2 &=& \mph \left(\frac{ \frac{1}{c \xph} - 1 }{\frac{1}{c x} -
    1}\right)^{-a} \nn\\
&=& \mph \lb \frac{ \frac{1}{c \xph} - 1 }{W \left( \left(
        \mph\right)^{\frac{1}{2 c}} \left( \frac{1}{c \xph} - 1
      \right) \e^{\frac{1}{c \xph} - 1} \right)} \rb^{-a}
\end{eqnarray}

\section{A trial effective potential consistent with RG\label{Veff}}

In \cite{Sonoda:2013a}, the effective potential for the large $N$
limit of the $O(N)$ linear sigma model in four dimensions has been
obtained as
\begin{equation}
\frac{d}{d \frac{v^2}{2}} \Veff (v) = \mph \left(x, m^2 + (4\pi)^2 x
  \frac{v^2}{2} \right)
\label{largeN}
\end{equation}
where $v$ is the VEV of the scalar field with no anomalous dimension:
\begin{equation}
\dt v = v
\end{equation}
In the large $N$ limit, we obtain
\begin{equation}
c = 0,\quad a = -1
\end{equation}
so that
\begin{equation}
\lb\begin{array}{r@{~=~}l}
\Lambda (x) & \e^{\frac{1}{x}}\\
\mph (x,m^2) & \Lambda (x)^2 \exp \left[ W_{-1} \left( - 2
    \frac{m^2}{x \Lambda (x)^2} \right) \right]
\end{array}\right.
\end{equation}
In the symmetric phase $m^2 > 0$, $\mph (x,m^2)$ gives the physical
squared mass of the scalar fields $\phi^I\,(I=1,\cdots,N)$.  In the
broken phase $m^2 < 0$, the physical squared mass vanishes at
\begin{equation}
v^2 = \frac{- m^2}{(4 \pi)^2 x}
\end{equation}

To generalize (\ref{largeN}) for a finite $N$, for which $c$ and $a$
are given by (\ref{c-ON}) and (\ref{a-ON}), we may try
\begin{equation}
\frac{d}{d \frac{v^2}{2}} \Veff (v)
= z (\xph') \mph \left( x, m^2 + g (\xph') \left(\frac{1}{x}-c\right)^a
  \frac{v^2}{2} \right) \ge 0 \label{trial} 
\end{equation}
which is a monotonically increasing function of $v^2$.  Here $z$ and
$g$ are positive functions of the RG invariant
\begin{equation}
\xph' \equiv \xph (x, |m^2|)
\end{equation}
which is well-defined irrespective of the sign of $m^2$.  Note that
the term added to $m^2$ satisfies the same RG equation as $m^2$:
\begin{equation}
\dt \left[ \left(\frac{1}{x}-c\right)^a \frac{v^2}{2} \right] = (2 + a
x) \left[ \left(\frac{1}{x}-c\right)^a \frac{v^2}{2} \right] 
\end{equation}
There is no justification for (\ref{trial}) except that it is fully
consistent with RG, and that it gives the correct result in the large
$N$ limit where $z$ and $g$ are mere constants:
\begin{equation}
z = 1,\quad g = (4 \pi)^2
\end{equation}
Note that in the broken phase $m^2 < 0$, the effective potential (or
equivalently (\ref{trial})) is defined only for
\begin{equation}
    v^2 \ge v_{\mathrm{min}}^2 
    \equiv \frac{- 2 m^2}{g (\xph')} \left(\frac{1}{x} - c\right)^{-a}
    > 0
\end{equation}
Since the right-hand side of (\ref{trial}) is monotonically increasing
with $v^2$, the effective potential is minimized at $v^2 =
v^2_{\mathrm{min}}$.

The main advantage of the assumption (\ref{trial}) is its
integrability.  To integrate (\ref{trial}) with respect to $v$, we use
(\ref{mph-expression}) to write (\ref{trial}) as
\begin{eqnarray}
\frac{d}{d \frac{v^2}{2}} \Veff (v) &=& z (\xph') 
\left( \frac{m^2}{\left(\frac{1}{x} -
      c\right)^a} + g (\xph') \frac{v^2}{2} \right) \nn\\
&&\quad \times \lb \frac{a+2c}{2}
W \left[ \frac{2 c^{- \frac{2c}{a+2c}}}{a + 2 c} \left(
      \frac{1}{\Lambda (x)^2}
\left( \frac{m^2}{\left(\frac{1}{x}-c\right)^a} + g (\xph')
  \frac{v^2}{2} \right) \right)^{- \frac{1}{a + 2 c}} \right] \rb^a
\end{eqnarray}
where $W$ should be $W_{-1}$ for $a + 2 c < 0$.  Denoting
\begin{equation}
\eta \equiv \frac{1}{\Lambda (x)^2} \left(
\frac{m^2}{\left(\frac{1}{x}-c\right)^a} + g (\xph') \frac{v^2}{2}
\right) \ll 1
\end{equation}
we can rewrite the differential equation for $\Veff$ as
\begin{equation}
\frac{d}{d \eta} \Veff = \frac{z (\xph')}{g (\xph')} \Lambda (x)^4\,
 \eta \lb \frac{a + 2 c}{2} W \left( \frac{2 c^{- \frac{2
         c}{a+2c}}}{a+2c} \eta^{- \frac{1}{a + 2 c}} \right) \rb^a
\end{equation}
We thus obtain
\begin{equation}
\Veff (v) = \frac{z (\xph')}{g (\xph')} \Lambda (x)^4 \int_0^\eta d\eta\,
\eta \lb \frac{a + 2 c}{2} W \left( \frac{2 c^{- \frac{2
         c}{a+2c}}}{a+2c} \eta^{- \frac{1}{a + 2 c}} \right) \rb^a
\end{equation}
Using the formulas
\begin{eqnarray}
&&\int_s^\infty ds\, s^{\beta-1} W(s)^\alpha \nn\\
&&\quad = - (-
\beta)^{-1-\alpha-\beta} \left[ \beta \Gamma \left(\alpha+\beta, -
        \beta W (s)\right) - \Gamma \left(\alpha+\beta+1, - \beta W
        (s)\right) \right]\nn\\
&&\quad (\beta < 0, s > 0)\\
&&\int_s^0 ds\, (-s)^{\beta - 1} (- W_{-1} (s))^\alpha \nn\\
&&\quad = -
\beta^{-1-\alpha-\beta} \left[ \beta \Gamma \left( \alpha+\beta, -
        \beta W_{-1} (s)\right) - \Gamma \left( \alpha+\beta+1, -
        \beta W_{-1} (s)\right) \right]\nn\\
&&\quad (\beta > 0, s < 0)
\end{eqnarray}
where
\begin{equation}
\Gamma (a,z) \equiv \int_z^\infty dt\, t^{a-1} \e^{-t}
\end{equation}
is the incomplete gamma function, we finally obtain
\begin{eqnarray}
&& \Veff (v) = \frac{z (\xph')}{g (\xph')} \,2^{2 a + 8 c -1}
c^{4 c}\, \Lambda (x)^4 \nn\\
&&\quad \times \Big\lbrace
2 (a + 2 c) \Gamma \left( - a - 4 c, 2(a+2 c) W (s)\right) + \Gamma
\left(-a - 4 c + 1, 2(a+2 c) W (s)\right) \Big\rbrace
\label{Veff-trial}
\end{eqnarray}
where 
\begin{equation}
s \equiv \frac{2}{a + 2 c} c^{- \frac{2c}{a+2c}} \eta^{- \frac{1}{a + 2
    c}}
\end{equation}
Note that $W$ should be replaced by $W_{-1}$ for $a + 2 c < 0$.

\section{Conclusions\label{conclusions}}

In this paper we have constructed two physical parameters $\xph$ and
$\mph$ by solving generic 2-loop RG equations analytically in terms of
the Lambert $W$ function.  In addition we have constructed explicitly
a trial effective action, which is fully consistent with RG, by
generalizing the analytic expression for the large $N$ limit of the
$O(N)$ linear sigma model in four dimensions.\cite{Sonoda:2013a} The
trial effective potential is, however, at best a wild guess at the
true effective potential.  Its only merit may be that it gives an
intriguing example of what RG improved perturbation theory can
produce.

The closed-form analytic expressions for $\xph$ (given by (\ref{xph}))
and $\mph$ (given by (\ref{mph})) sum the corresponding perturbative
series.  Further studies may elucidate the precise asymptotic nature
of the perturbative expansions, as has been done for
QCD.\cite{Gardi:1998qr, Magradze:1998ng}

\appendix
\section{The Lambert $W$ function}

\begin{figure}[h]
\includegraphics[width=8cm]{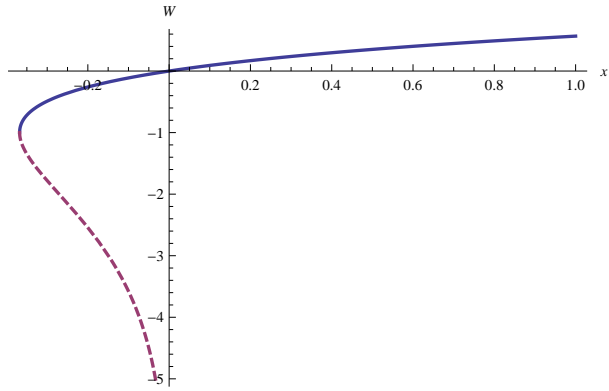}
\caption{The real valued Lambert $W$ function has two branches: the
  upper $W_0$ (solid) and lower $W_{-1}$ (dashed).}
\label{lambert_w}
\end{figure}

The Lambert $W$ function is defined implicitly by
\begin{equation}
W (x) \e^{W (x)} = x
\end{equation}
or equivalently by
\begin{equation}
W (x \e^x) = x
\end{equation}
Restricted to real values, the function has two branches: the upper
$W_0 (x) > -1$ defined for $x \in [- \e^{-1}, +\infty)$ and the lower
$W_{-1} (x) < -1$ for $x \in [- \e^{-1}, 0)$.  (See
Fig.~\ref{lambert_w}.)  For simplicity, we denote $W_0$ as $W$ in this
paper. 
We obtain the following asymptotic expansions:
\begin{enumerate}
\item For $x \gg 1$,
\begin{equation}
W_0 (x) = \ln x - \ln \ln x + \mathrm{O} \left( \frac{\ln \ln x}{\ln
    x} \right)
\end{equation}
\item For $- x \ll 1$, 
\begin{equation}
W_{-1} (x) = \ln (-x) - \ln \left(- \ln (-x)\right) + \mathrm{O}
\left(\frac{\ln \left(- \ln (-x)\right)}{\ln (-x)}\right)
\end{equation}
\end{enumerate}

\section{Asymptotic free theories}

Let us quickly summarize the applications of the Lambert $W$ function
to asymptotic free theories.\cite{Gardi:1998qr, Magradze:1998ng} For
asymptotic free theories, the generic 2-loop RG equation is
\begin{equation}
\dt x = x^2 + c x^3
\end{equation}
For example, in QCD with $n_f$ flavors, we find
\begin{equation}
c = \frac{6 \left(153 - 19 n_f\right)}{33 - 2 n_f}
\end{equation}
and in the $O(N)$ non-linear sigma model in two dimensions, we find
\begin{equation}
c = \frac{1}{N-2}
\end{equation}

The scale parameter in this case is defined by
\begin{equation}
\Lambda (x) \equiv \left(\e^{- \frac{1}{c x} - 1} \left(\frac{1}{c
      x} + 1\right)\right)^c 
\end{equation}
Inverting this, we obtain
\begin{equation}
x = \frac{1}{- c \lb 1 + W \left( - \left( \Lambda
      (x)\right)^{\frac{1}{c}} \right) \rb}
\end{equation}
Hence, the running parameter is given by
\begin{equation}
\xbar (t; x) = \frac{1}{- c \lb 1 + W \left( - \left( \e^t
      \Lambda
      (x)\right)^{\frac{1}{c}} \right) \rb}
\end{equation}
which satisfies
\begin{equation}
\partial_t \xbar (t; x) = \xbar (t;x)^2 + c \xbar (t;x)^3
\end{equation}
and $\xbar (0; x) = x$.

The large $t$ behavior of $\xbar (t;x)$ depends on the sign of $c$:
\begin{itemize}
\item For $c > 0$, $\xbar (t;x)$ diverges as $t \to t_{\mathrm{max}}$, where
$t_{\mathrm{max}}$ is given by
\begin{equation}
\e^{- t_{\mathrm{max}}}  = \e^{c}\, \Lambda (x)
\end{equation}
\item For $c < 0$, $\xbar (t;x)$ approaches $\frac{1}{-c}$ as $t \to +
\infty$.
\end{itemize}


\bibliography{lambertRG-v2}

\end{document}